\documentclass[12pt,english]{article}
\usepackage{lscape}
\usepackage[english]{babel}

\usepackage{eurosym} 
\usepackage{comment}
\usepackage{soul}

\usepackage[a4paper,
            left=1in,
            right=1in,
            top=1in,
            bottom=1in]{geometry}

\usepackage{amsmath}
\usepackage{graphicx}
\usepackage[colorlinks=true, allcolors=blue,hyperfootnotes=false]{hyperref}
\usepackage{tabularx}
\usepackage{booktabs}
\usepackage{adjustbox}
\usepackage{amssymb}
\usepackage{amsfonts}
\usepackage{placeins}
\usepackage{dsfont}
\usepackage{xcolor}
\usepackage{soul}
\usepackage[comma,authoryear]{natbib}
\usepackage{multibib}
\newcites{sec}{References in the Online Appendix}

\usepackage{siunitx}

\setcitestyle{citesep={,}}

\usepackage{caption}
\usepackage{subcaption}
\usepackage[flushleft]{threeparttable}

\usepackage{setspace}

    \usepackage[toc,page,header]{appendix}
    \usepackage{minitoc}
    \noptcrule 

\allowdisplaybreaks

\newtheorem{assumption}{Assumption}

\newtheorem{proposition}{Proposition}

\AddToHook{cmd/appendix/before}{%
  \setcounter{equation}{0}%
  \renewcommand{\theequation}{\thesection.\arabic{equation}}%
  \setcounter{theorem}{0}%
  \setcounter{proposition}{0}
}

\begin{document}

\title{\LARGE{Concentration and Markups in International Trade}\thanks{We thank Laura Castillo-Martinez, Sebastian Heise, and Yannis Papadakis for insightful discussions. We are also grateful to Rodrigo Adao, Ariel Burstein, Thomas Chaney, Jan De Loecker, Michael Peters, Esteban Rossi-Hansberg, and many seminar and conference participants for helpful feedback. Andrew Sharng and Nicolas Wesseler provided excellent research assistance. All errors are our own.}}
\author{\normalsize Vanessa Alviarez \quad\quad Michele Fioretti\quad\quad Ken Kikkawa \quad\quad Monica Morlacco\thanks{Alviarez: IADB, valviarezr@iadb.org. 
Fioretti: Bocconi University, michele.fioretti@unibocconi.edu. Kikkawa: UBC Sauder and NBER, ken.kikkawa@sauder.ubc.ca. Morlacco, USC, morlacco@usc.edu.}}

\date{July 31, 2025}

\maketitle

\begin{abstract}
This paper derives a closed-form expression linking aggregate markups on imported inputs to concentration in a model of firm-to-firm trade with two-sided market power. Our theory extends standard oligopoly insights in two dimensions. First, it reveals that markups increase with exporter concentration and decrease with importer concentration, reflecting the balance of oligopoly and oligopsony forces. Second, it adapts conventional market definitions to reflect rigid trading relationships, yielding new concentration measures that capture competition in firm-to-firm trade. Analysis of Colombian transaction-level import data shows these differences are key to understanding markup dynamics in international trade.

\vspace{7cm}
\end{abstract}
\pagebreak{}

\section{Introduction\label{sec:Introduction}}
\doublespacing
The expansion of global supply chains has placed intermediate inputs at the center of economic activity. Accounting for a large share of gross output and international trade, these inputs are critical to understanding production costs and consumer prices. Markets for intermediate goods often function as \emph{bilateral oligopolies}, where a small number of large exporters and importers repeatedly negotiate prices in environments marked by high entry and switching costs \citep{antras2015global, alviarez2024twosided}. Such dominance by a few global firms raises questions about the evolution of market power in input trade.

Because markups are challenging to observe directly, concentration is often used in competition analysis as a proxy for pricing power \citep{nocke2022concentration}. This proxy is known to be noisy, particularly in environments that deviate from textbook oligopoly models \citep{syverson2019macroeconomics}. However, in bilateral oligopoly settings such as intermediate input markets, concentration can still provide valuable information about the exercise of market power \citep{hendricks2010theory}.

This paper shows that concentration can be informative about markups in input trade, provided that both sides of the market and network rigidities are considered. Building on the two-sided bargaining model of \citet{alviarez2024twosided}, our first contribution is to derive a closed-form expression linking aggregate markups on imported inputs to exporter and importer concentration. Markups rise with exporter concentration, reflecting oligopoly power, and fall with importer concentration, reflecting countervailing oligopsony power. The strength of each force depends on supply and demand elasticities, while their net effect is shaped by relative bargaining power.

Our second contribution is to clarify how network rigidities affect the measurement of concentration. Much of the literature defines markets at broad industry levels, assuming all firms compete with all others within a sector. When switching is costly and relationships are persistent, however, firms interact with only a limited set of partners. To capture this, our theory introduces two \emph{network-based} concentration indices tailored to this setting: an exporter-side index, which aggregates the Herfindahl–Hirschman Indices (HHIs) of each importer’s supplier base, and an importer-side index, defined analogously. These measures nest standard industry-level HHIs as special cases, revealing that they can substantially understate true concentration.

We assess the empirical relevance of our measures using Colombian trade data from 2011 to 2020. Our network-based indices show a steady rise in average exporter concentration and relatively stable importer concentration. In contrast, standard industry-level HHIs exhibit volatile exporter concentration and a marked decline in importer concentration. These discrepancies suggest that conventional measures not only understate the extent of concentration in firm-to-firm trade but also misrepresent its evolution over time.

These differences matter for markups. We estimate that bargaining power is relatively balanced between Colombian importers and foreign exporters, underscoring the empirical relevance of bilateral market power. Combined with concentration trends, these estimates point to rising markups over time, driven mainly by increasing exporter concentration, with importer concentration providing a partial offset. By contrast, standard approaches that focus only on suppliers and rely on industry-wide concentration measures imply more volatile but largely flat markup trends. Our findings demonstrate that accounting for both bargaining power and network rigidities is essential for accurately capturing the dynamics of markups in firm-to-firm trade.

Two caveats are worth noting. First, our analysis focuses on input markups. While upstream markups have clear welfare implications, as illustrated in a simplified extension, evaluating these effects requires additional data on downstream demand, pass-through, and market structure, which are not observable in our trade data. Second, the empirical analysis relies on simplifying assumptions introduced to align the model with available data. These assumptions are not inherent to the theory and could be relaxed with richer data in future work.

Second, while the model implies a mapping from concentration measures to markups, this mapping relies on structural parameters such as demand and supply elasticities and bargaining power, which are assumed to remain stable.  However, concentration is itself an equilibrium outcome that may reflect underlying shifts in economic fundamentals. For this reason, caution is warranted when interpreting the observed trends, especially when the forces driving changes in concentration over time are not well understood.

This paper contributes to a growing literature on the relationship between concentration and market power, with a focus on intermediate input markets. A large body of work, primarily using U.S. data, documents two concurrent trends: a sustained rise in industry concentration \citep[e.g.,][]{grullon2019us, koltay2023concentration, kwon2024100} and an increase in average markups \citep[e.g.,][]{de2020rise, de2018global, hall2018new}. These trends are often viewed as indicative of growing market power, although some have argued they instead reflect technological change and increasing returns to scale.\footnote{See, e.g., \citet{autor2020fall, covarrubias2020good, traina2018}. For critical reviews, see \citet{syverson2019macroeconomics} and \citet{eeckhout2021book}.}

Our paper contributes to this debate by showing how concentration and markups are linked through production networks. Building on work that relates markups to industry-level HHIs in oligopoly models \citep[e.g.,][]{Grassi2018, burstein2025bottom, nocke2025concentration}, we extend this logic to a bargaining setting with market power on both sides. We show that two-sided concentration acts as a sufficient statistic for aggregate markups, providing a theoretical basis for using concentration measures as proxies for market power in firm-to-firm trade, while highlighting the role of network structure and trading frictions.\footnote{Recent work underscores why such frictions matter: \cite{martin2023relationship} shows that firm‑to‑firm trade links are highly sticky and that this amplifies the response of trade flows to uncertainty shocks, while \cite{kikkawa2022imperfect} and \cite{alviarez2024twosided} show that these rigidities shape prices and tariff pass‑through.}

We also contribute to the debate over market definition in concentration analysis. Prior work shows that measured concentration is highly sensitive to how markets are defined: using local rather than national market definitions can reverse trends \citep{rossi2021diverging}, and accounting for import competition can flatten them \citep{amitiheise2024us}.\footnote{See also \citet{pellegrino2025product} on product similarity and \citet{chen2016industry} on the manipulability of industry classifications.} We focus on the role of network rigidities: when firms cannot easily switch partners, competition occurs along existing links. These rigidities create pair-level markets, and we show they shape both the level and evolution of concentration.

The closest study is \citet{hendricks2010theory}, which also stresses the link between buyer and supplier concentration and markups, but in a homogeneous‑goods setting. Our framework generalizes theirs by accommodating firm heterogeneity and arbitrary network structure. Moreover, instead of relying on a stylized supply‑function game, we derive markups from a Nash-in-Nash bargaining model. Our approach highlights how network rigidities define effective market boundaries and concentration, and it is tractable enough to be implemented with standard customs data. These features make it well-suited for applied work on market power in global production networks.

\section{Theory\label{sec:Theory}}

We study a firm-to-firm trade environment with two-sided market power. Sections \ref{subsec:Environment} and \ref{subsec:Bargaining} introduce the environment and characterize equilibrium at the pair level, building on the framework in \citet{alviarez2024twosided}. We summarize only the elements needed to derive our main result on aggregate markups, and refer to the companion paper for a full exposition. Section \ref{subsec:Aggregate-Markups} describes the aggregation and presents our main theoretical results.

\subsection{Environment \label{subsec:Environment}}

The industry consists of a finite number of exporters (indexed by \(i\)) and importers (indexed by \(j\)) engaged in the trade of differentiated intermediate inputs. Each importer \(j\) is matched with a fixed set of exporters \((\mathcal{Z}_j)\), and each exporter \(i\) supplies a fixed set of importers \((\mathcal{Z}_i)\). These firm-to-firm relationships are taken as given.

Each exporter \(i\) produces under decreasing returns to scale, with marginal cost increasing in total output \(q_i\):
\begin{equation*}
    c_i = c_i(q_i) = k_i q_i^{\frac{1-\theta}{\theta}},
    \label{eq:c_i}
\end{equation*}
where \(k_i\) reflects productivity or upstream input costs, and \(\theta \in (0,1]\) governs the degree of returns to scale. A lower \(\theta\) implies steeper marginal cost curves. When \(\theta = 1\), the technology exhibits constant returns, and the marginal cost is constant.

Each importer \(j\) combines input varieties from its matched exporters using a CES aggregator with elasticity of substitution \(\rho > 1\). The resulting composite foreign input enters a Cobb-Douglas production function for the final good. Production follows a nested CES structure:
\begin{equation}
    q_j = \varphi_j \left(\vphantom{q_j^f}q_j^d\right)^{1-\gamma} \left(q_j^f\right)^\gamma 
    \quad \text{with} \quad 
    q_j^f = \left(\sum_{i \in \mathcal{Z}_j} q_{ij}^{\frac{\rho - 1}{\rho}}\right)^{\frac{\rho}{\rho - 1}},
    \label{eq:q_j}
\end{equation}
where \(q_j^d\) denotes domestic inputs, \(\varphi_j\) is importer \(j\)’s productivity, and \(\gamma \in (0,1)\) denotes both the output elasticity and variable cost share of the foreign input.\footnote{The assumptions of constant returns to scale and constant demand shifters across exporters' varieties are without loss of generality; none of the results hinge on these simplifications.} 
The final good \(q_j\) is sold in monopolistically competitive markets with CES demand and elasticity \(\nu > 1\).

\subsection{Bargaining and Pair-Level Markups \label{subsec:Bargaining}}

Negotiations follow a two-stage protocol. In the first stage, each importer chooses the quantity \(q_{ij}\) of each matched exporter variety to minimize variable costs, taking the price \(p_{ij}\) as given. This yields the importer’s demand function:
\begin{equation}
q_{ij} = c_j \, q_j\, p_{ij}^{-\rho} \left(p_j^f\right)^{\rho - 1},
\label{eq:demand}
\end{equation}
where \(c_j\) is importer \(j\)’s marginal cost, \(q_j\) is its total output, and \(p_j^f = \left( \sum_{i} \  p_{ij}^{1 - \rho} \right)^{1/(1 - \rho)}\) is the CES price index for the foreign input bundle.

In the second stage, each importer-exporter pair jointly negotiates the bilateral price \(p_{ij}\), taking as given the demand function in equation \eqref{eq:demand}. The bilateral price solves:
\begin{equation}
\max_{p_{ij}} \left( \underbrace{\pi^{i} - \tilde{\pi}^{i}_{(-j)}}_{GFT_{ij}^{i}} \right)^{1 - \phi}
\left( \underbrace{\pi^{j} - \tilde{\pi}^{j}_{(-i)}}_{GFT_{ij}^{j}} \right)^{\phi},
\label{eq:nash_bargaining}
\end{equation}
where \(\phi \in (0,1)\) denotes the importer's bargaining weight. The negotiated price thus splits the \emph{gains from trade} between the two parties, \(GFT_{ij}^k\) for \(k \in \{i,j\}\), defined as the increase in profits from participating in the match relative to the outside option in which the match is severed but all other relationships remain unchanged.

To solve problem~\eqref{eq:nash_bargaining}, we adopt the Nash-in-Nash solution concept \citep{Horn1988}, whereby each bilateral negotiation treats the outcomes of all other relationships in the network as fixed. Accordingly, we omit explicit dependence on prices and quantities in other matches throughout the analysis.

\paragraph{Equilibrium Markups}
The solution to the bargaining problem yields a price equal to a markup over marginal cost, \(p_{ij} = \mu_{ij} \, c_i\).  The equilibrium markup is a weighted average of the oligopoly (\(\phi=0\)) and oligopsony (\(\phi=1\)) benchmarks,
\begin{equation}
\mu_{ij} = (1-\omega_{ij})\,\mu^{\text{oligopoly}}(s_{ij}) + \omega_{ij}\,\mu^{\text{oligopsony}}(x_{ij}),
\label{eq:markup}
\end{equation}
where \(\omega_{ij}\) is the effective bargaining weight, as explained below.  Because bargaining is bilateral and occurs only within existing supplier–buyer matches, the relevant market for each transaction is effectively defined at the pair level: competition is confined to importer \(j\)’s current suppliers in $\mathcal{Z}_j$ and exporter \(i\)’s current buyers in $\mathcal{Z}_i$.  Accordingly, both components and weights in equation~\eqref{eq:markup} depend on two \emph{bilateral market shares},
\begin{equation}
s_{ij} \equiv \frac{p_{ij} q_{ij}}{\sum_{k \in \mathcal{Z}_j} p_{kj} q_{kj}}
    \quad\quad\text{and}\quad\quad
x_{ij} \equiv \frac{q_{ij}}{\sum_{k \in \mathcal{Z}_i} q_{ik}},
\label{eq:s_and_x_def}
\end{equation}
where \(s_{ij}\) is exporter \(i\)’s \emph{supplier share} in importer \(j\)’s total input expenditure and \(x_{ij}\) is importer \(j\)’s \emph{buyer share} of exporter \(i\)’s total output. These shares will form the basis for our \emph{network-based} concentration measures later on.

The \emph{oligopoly markup} applies when the importer has no bargaining power (\(\phi = 0\)) and is given by:
\begin{equation}
\mu^{\text{oligopoly}}(s_{ij}) = \frac{\varepsilon_{ij}}{\varepsilon_{ij} - 1}, 
\quad \text{with} \quad 
\varepsilon_{ij} = \rho \ (1 - s_{ij}) + \eta \ s_{ij},
\label{eq:mu_oligopoly}
\end{equation}
where \(\varepsilon_{ij} \equiv -\frac{d\ln q_{ij}}{d\ln p_{ij}}\) is the residual demand elasticity faced by exporter \(i\), and \(\eta \equiv 1 - \gamma + \nu \gamma\) reflects the elasticity of the importer’s foreign input bundle ($q_j^f$) with respect to its price index ($p_j^f$). Provided \(\rho > \eta\), the markup increases with the supplier share \(s_{ij}\), as larger suppliers face less elastic demand. This corresponds to standard oligopoly models in trade, such as \citet{Atkeson2008} and \citet{kikkawa2022imperfect}.

The \emph{oligopsony markdown} applies when the importer has full bargaining power ($\phi = 1$) and is given by:
\[
\mu^{\text{oligopsony}}(x_{ij}) = \theta \cdot \frac{1 - (1 - x_{ij})^{1/\theta}}{x_{ij}} \in [\theta, 1].
\]
It declines with the buyer share $x_{ij}$, reflecting the importer’s ability to extract surplus. Under decreasing returns ($\theta < 1$), marginal cost exceeds average cost, so large buyers can negotiate prices below cost by capturing quasi-rents. When $\theta = 1$, marginal and average cost coincide, no rents are available, and the markdown equals one regardless of $x_{ij}$.

The weight \(\omega_{ij} \in (0,1)\) can be interpreted as the importer’s {effective} bargaining power:
\begin{equation}
\omega_{ij} = \frac{\frac{\phi}{1 - \phi} \lambda_{ij}}{1 + \frac{\phi}{1 - \phi} \lambda_{ij}}
\quad \quad \text{with} \quad \quad
\lambda_{ij} = \frac{s_{ij}}{(1-(1-s_{ij})^{\frac{\eta-1}{\rho-1}})} \cdot \frac{\eta - 1}{\varepsilon_{ij} - 1} \geq 1. \label{weight}
\end{equation}

The parameter \(\phi\) reflects baseline importer bargaining power, while \(\lambda_{ij}\) scales it based on the match’s importance within a buyer's network, as captured by the supplier share $s_{ij}$. This term is a hump-shaped function of the supplier share \(s_{ij}\), with \(\lambda_{ij} \to 1\) (and thus \(\omega_{ij} \to \phi\)) as \(s_{ij} \to 0\) or \(s_{ij} \to 1\). Higher \(\lambda_{ij}\) means higher leverage for importer \(j\), increasing \(\omega_{ij}\) and shifting prices toward the oligopsony limit.

\subsection{Aggregate Markups\label{subsec:Aggregate-Markups}}

We define the aggregate markup on imported inputs as the ratio of total sales to total variable costs. Under this definition, the aggregate markup is exactly equal to the harmonic mean of bilateral markups, weighted by each match’s share of total expenditure:
\begin{equation}
\mu \equiv \frac{\sum_{i} \sum_{j} p_{ij} q_{ij}}{\sum_{i} \sum_{j} \theta c_i q_{ij}} = \left( \sum_{i} \sum_{j} \iota_{ij} \mu_{ij}^{-1} \right)^{-1}, \label{eq:exact_agg_mkup}
\end{equation}
where $\iota_{ij} \equiv \frac{p_{ij} q_{ij}}{\sum_{i} \sum_{j} p_{ij} q_{ij}}$.

To tractably approximate equation~\eqref{eq:exact_agg_mkup}, we impose two simplifying assumptions:

\begin{assumption}\label{A1}
    The bargaining weight is approximately constant across all matches, i.e., \( \omega_{ij} \approx \phi \) for all \( (i,j) \).
\end{assumption}

This approximation becomes exact in the limits \( s_{ij} \to 0 \) and \( s_{ij} \to 1 \), and is satisfied when the distribution of \( \lambda_{ij} \) is tightly centered around one in the data.

\begin{assumption}\label{A2}
    The dispersion of bilateral markups \( \mu_{ij} \) is sufficiently limited to support the approximation of the harmonic mean by the arithmetic mean. 
\end{assumption}

Section \ref{sec:Identification-and-estimation} shows that both assumptions are supported under our calibrated parameters.

Under Assumptions \ref{A1} and \ref{A2}, the aggregate markup admits a closed-form approximation as a function of average concentration on both sides of the buyer--supplier network.

Specifically, the approximation yields two concentration indices tailored to our trading environment. To capture supplier-side (exporter) concentration, we compute for each importer $j$ the HHI of its supplier shares, defined as \(\text{HHI}^{\text{s}}_j = \sum_i s_{ij}^2\). We then aggregate these across importers using each importer's share of total industry trade as weights, i.e.:
\begin{equation}
HHI^{\text{suppliers}} \equiv \sum_j \varphi_j\,\text{HHI}^{\text{s}}_j,
\label{eq:HHI_suppliers}
\end{equation}
where \(\varphi_j = \frac{\sum_i p_{ij} q_{ij}}{\sum_{i,j} p_{ij} q_{ij}}\). This index summarizes the overall degree of exporter concentration in the industry.

To capture buyer-side (importer) concentration, we apply a parallel logic. For each exporter \(i\), we define a modified HHI of its buyers that combines revenue and quantity shares \(\text{MHHI}^{x}_i = \sum_j x_{ij}^r x_{ij},\) where \(x_{ij}^r = \frac{p_{ij} q_{ij}}{\sum_k p_{ik} q_{ik}}\) is the revenue share of importer \(j\) in exporter \(i\)’s total sales, and \(x_{ij}\) is the quantity-based buyer share defined in equation~\eqref{eq:s_and_x_def}. We then average these exporter-level indices using each exporter's share of total industry trade as weights, i.e.:
\begin{equation}
HHI^{\text{buyers}} \equiv \sum_{i} \varphi_i\, \text{MHHI}^{x}_i,
\label{eq:HHI_buyers}
\end{equation}
where \(\varphi_i = \frac{\sum_j p_{ij} q_{ij}}{\sum_{i,j} p_{ij} q_{ij}}\). The inner sum, \(\text{MHHI}^{x}_i\), captures how concentrated each exporter’s sales are across buyers. Unlike a standard HHI, it combines a revenue share and a quantity share. This asymmetry reflects the fact that, for buyer-side markdowns, the relevant notion of share is quantity-based.

We can now state the main result of this section.

\begin{proposition}\label{prop:1}
\textit{Under Assumptions \ref{A1} and \ref{A2}, the aggregate markup defined in equation \eqref{eq:exact_agg_mkup} admits the following approximation:}
\begin{align}
\mu 
\approx & \left(1 - \phi\right)\frac{\rho}{\rho - 1} + \phi \nonumber \\
& + \left(1 - \phi\right)\left[\frac{\rho - \eta}{\left(\rho - 1\right)^2} HHI^{\text{suppliers}}\right] \nonumber \\
& - \phi \left[\frac{1 - \theta}{2\theta} HHI^{\text{buyers}}\right],\label{eq:agg_mkup}
\end{align}
where $HHI^{\text{suppliers}}$ and $HHI^{\text{buyers}}$ are defined in equations~\eqref{eq:HHI_suppliers} and~\eqref{eq:HHI_buyers}, respectively. 
\end{proposition}

\noindent\textbf{Proof.} See Appendix~\ref{subsec:Derivation-of-approximation}.

\vspace{1em}
Proposition~\ref{prop:1} establishes that, under suitable regularity conditions, the aggregate markup can be approximated by a linear function of concentration on both sides of the market. %

Under minimal concentration ($HHI^{\text{suppliers}} \to 0$, $HHI^{\text{buyers}} \to 0$), the aggregate markup collapses to a convex combination of the monopolistic competition case \(\left(\frac{\rho}{\rho - 1}\right)\) and the monopsonistic competition case \((1)\), with the bargaining parameter $\phi$ as the weight. 

An increase in exporter concentration, as measured by $HHI^{\text{suppliers}}$, raises the aggregate markup, reflecting greater oligopoly power. The magnitude of this effect declines with the elasticity of substitution $\rho$: greater substitutability among suppliers weakens their ability to raise prices. The effect vanishes in the limits $\rho \to \infty$ (perfect competition among suppliers) or $\phi \to 1$ (importers fully dictate prices).

In contrast, an increase in importer concentration, as measured by $HHI^{\text{buyers}}$, reduces the aggregate markup, capturing the influence of countervailing oligopsony power. This effect diminishes with the returns-to-scale parameter $\theta$, as flatter marginal cost curves limit buyers' ability to depress prices. The influence of importer concentration disappears in the limit $\theta \to 1$ (perfectly elastic supply) or $\phi \to 0$ (exporters fully dictate prices).

\subsubsection{Comparison with Standard Approaches}

Proposition~\ref{prop:1}  generalizes the well-known link between concentration and markups in differentiated-product oligopoly models \citep[e.g.,][]{Grassi2018, burstein2025bottom, nocke2025concentration} to a bargaining setting with bilateral market power.

This generalization relies on two key departures. First, prices are set bilaterally rather than unilaterally by suppliers. When buyers lack bargaining power ($\phi \to 0$), pricing is only governed by supplier concentration, as in standard oligopoly models. When buyers have bargaining power ($\phi > 0$), buyer concentration also plays a role. In the limit of no supplier power ($\phi \to 1$), markups are fully determined by buyer concentration, paralleling standard oligopsony models \citep[e.g.,][]{berger2022labor, amodio25}.\footnote{Although the $\phi \to 1$ case differs in structure from traditional oligopsony models, where quantities appear on the supply side and markdowns are defined as the wedge between input prices and marginal revenue products, the core logic is similar. In our framework, quantities remain on the demand side and markdowns are defined as price--cost ratios. See \citet{alviarez2024twosided} for a detailed comparison.}

Second, the relevant market is defined by the set of active links, so that market shares are given by equation~\eqref{eq:s_and_x_def}. This gives rise to the exporter and importer concentration indices in equations~\eqref{eq:HHI_suppliers} and~\eqref{eq:HHI_buyers}, which we refer to as \emph{network-based} concentration indices.

By contrast, standard models assume industry-wide competition on each side of the market, leading to conventional HHIs based on total trade shares:
\begin{align}
HHI^{\text{suppliers, std}} &\equiv \sum_{i} \varphi_{i}^{2}, \label{eq:HHI_suppliers_std} \\
HHI^{\text{buyers, std}} &\equiv \sum_{j} \varphi_{j}^{2}, \label{eq:HHI_buyers_std}
\end{align}
where $\varphi_i$ and $\varphi_j$, defined in equations~\eqref{eq:HHI_suppliers} and~\eqref{eq:HHI_buyers}, denote exporter and importer shares in total trade. These industry-wide HHIs match the network-based measures only when all trade is concentrated in a single exporter or importer.

\subsubsection{Concentration and Consumer Prices}

Concentration and markups in input trade matter because they can affect consumer prices. Higher input markups raise marginal costs, which in turn raise final good prices.

Appendix~\ref{app:consumer-prices} formalizes this connection by extending the framework to show how upstream markups affect consumer prices through firms’ cost structures. Assuming a common cost share across firms, the consumer price index embeds the aggregate markup in equation~\eqref{eq:agg_mkup} and scales proportionally with it.

While this highlights the broader welfare relevance of our results, empirically pursuing this channel would require strong assumptions on technology and pass-through, along with data on final prices and domestic input costs--none of which are available. We therefore focus on imported input markups for the rest of our analysis.

\section{Empirical Analysis\label{sec:data}}

This section and the next apply our framework to firm-to-firm trade data. Our goal is not to test the model, as its aggregate implications follow directly from the underlying bargaining theory, already validated in \citet{alviarez2024twosided} using U.S. data.\footnote{See also \citet{gopinath2011search}, \citet{grossman2020tariffs}, \citet{atkin2024trade}, and \citet{cristoforoni2025oligopolies}.} Instead, we assess the quantitative importance of two-sided market power and network-based concentration measures in determining imported input markups in a new empirical setting.

We begin by describing the data and construction of key variables, then document how concentration evolves across markets. Section~\ref{sec:main_analysis} turns to markup implications.

\subsection{Data}

Our primary data source is the universe of Colombian customs records, covering all import transactions from 2011 to 2020.\footnote{Raw data are available at https://www.dian.gov.co/.} Each transaction reports the importer’s tax ID (NIT), the exporter’s name, the 10-digit HS product code, value, quantity, and other details. We aggregate the data at the importer-exporter-product-year level.

Importers are uniquely identified by NIT, while exporters are listed by name. To address variation due to accents, typos, and formatting, we apply machine learning–based cleaning to construct consistent supplier identifiers within each year.

Appendix Table~\ref{tab:Colombian-Imports:-Statistics} summarizes the data: on average, 2.5 million transactions per year, 147,600 buyer-supplier pairs, and USD 43.5 billion in trade. Comparisons with UN Comtrade confirm near-complete coverage of Colombia’s imports during this period.

\subsection{Measuring Key Variables of Interest\label{subsec:Measuring-key-variables}}

We make several adjustments to bring the model to the data. 

The model assumes a single foreign input $q_j^f$, but in practice, importers source multiple inputs from various foreign suppliers. To align with this data feature, we redefine $q_j^f$ in equation~\eqref{eq:q_j} as a Cobb--Douglas aggregator across HS10 inputs, each modeled as a CES composite of supplier varieties:
\begin{equation}
q_j^f = \prod_h \left(q_j^{f,h}\right)^{\alpha_h}, \quad \text{with} \quad q_j^{f,h} = \left(\sum_{i \in \mathcal{Z}^h_j} \left(q_{ij}^h\right)^{\frac{\rho - 1}{\rho}}\right)^{\frac{\rho}{\rho - 1}}, \label{eq:foreign_input_bundle}
\end{equation}
where  $\alpha_h = \frac{\sum_i \sum_j p_{ij}^h q_{ij}^h}{\sum_k \left( \sum_i \sum_j p_{ij}^k q_{ij}^k \right)}$ is the input’s expenditure share in total foreign inputs.

Two additional assumptions are needed to address data limitations. First, we assume that each importer $j$ sources product $h$ from either domestic or foreign suppliers, but not both.\footnote{Domestic inputs are captured in $q_j^d$ in equation~\eqref{eq:q_j}.} Thus, the sets $\{\mathcal{Z}_j^h\}_j$ in equation~\eqref{eq:foreign_input_bundle} include only foreign suppliers.

Under this assumption, the supplier share of exporter $i$ in importer $j$’s sourcing of product $h$ is:
\[
s_{ij}^{h} = \frac{p_{ij}^{h} \, q_{ij}^{h}}{\sum_{k \in \mathcal{Z}_{j}^{h}} p_{kj}^{h} \, q_{kj}^{h}}.
\]

Second, we assume destination-specific technologies, restricting the exporter’s relevant network to Colombian buyers. This reflects the fact that our data only includes shipments to Colombia. 

It follows that the buyer share of importer $j$ in exporter $i$’s Colombian sales of product $h$ is:
\begin{equation}
x_{ij}^{h} = \frac{q_{ij}^{h}}{\sum_{k \in \mathcal{Z}_{i}^{h, \text{COL}}} q_{ik}^{h}}, 
\quad \text{and} \quad 
x_{ij}^{r,h} = \frac{p_{ij}^{h} \, q_{ij}^{h}}{\sum_{k \in \mathcal{Z}_{i}^{h, \text{COL}}} p_{ik}^{h} \, q_{ik}^{h}}, \label{x}
\end{equation}
where $\mathcal{Z}_{i}^{h, \text{COL}}$ is the set of Colombian importers of product $h$ served by exporter $i$.

With this structure and definitions, and following the same logic used to derive equation~\eqref{eq:agg_mkup}, we substitute product-level markups \( \mu_h \) to express the aggregate input markup as a function of average concentration across foreign input products:
\begin{align}
\mu 
\approx & \left(1 - \phi\right)\frac{\rho}{\rho - 1} + \phi \nonumber \\
& + \left(1 - \phi\right)\left[\frac{\rho - \eta}{\left(\rho - 1\right)^2} \, \overline{HHI}^{\text{suppliers}}\right] \nonumber \\
& - \phi \left[\frac{1 - \theta}{2\theta} \, \overline{HHI}^{\text{buyers}}\right], \label{eq:agg_mkup_conc}
\end{align}
where
\begin{equation}
\overline{HHI}^{\text{suppliers}} \equiv \sum_h \alpha_h \, HHI_h^{\text{suppliers}}, 
\quad \text{and} \quad 
\overline{HHI}^{\text{buyers}} \equiv \sum_h \alpha_h \, HHI_h^{\text{buyers}}  \label{eq:agg_conc}
\end{equation}

are import-share-weighted average concentration indices across products, with product-level indices \( HHI_h^{\text{suppliers}} \) and \( HHI_h^{\text{buyers}} \) defined in equations~\eqref{eq:HHI_suppliers} and~\eqref{eq:HHI_buyers}, respectively.

For comparison, we also compute standard concentration indices based on industry-wide market shares. We define exporter and importer market shares at the product level as:
\[
\varphi_{it}^{h} \equiv \frac{\sum_{j \in \mathcal{Z}_{it}^{h, \text{COL}}} p_{ijt}^{h} q_{ijt}^{h}}{\sum_{i} \sum_{j} p_{ijt}^{h} q_{ijt}^{h}} \quad \text{and} \quad 
\varphi_{jt}^{h} \equiv \frac{\sum_{i \in \mathcal{Z}_{jt}^{h}} p_{ijt}^{h} q_{ijt}^{h}}{\sum_{i} \sum_{j} p_{ijt}^{h} q_{ijt}^{h}},
\]
respectively. Using these, we compute standard exporter and importer HHIs as defined in equations~\eqref{eq:HHI_suppliers_std} and~\eqref{eq:HHI_buyers_std}. Aggregate (standard) concentration measures are then obtained by taking import-share-weighted averages across products.

\subsection{Concentration Patterns Across Products}

Table~\ref{stats} presents summary statistics on exporter and importer concentration at the HS10 product-year level. A few patterns stand out.

\begin{table}[t]
\caption{Exporter and Importer Concentration: Summary Statistics\label{stats}}

\begin{centering}
\small
\begin{tabular}{lccccc}
\multicolumn{6}{c}{(a) Exporter Concentration} \\
\toprule
 & Mean & St. Dev. & p10 & p50 & p90 \\
\midrule
Nr. exporters (HS10) & 77.91 & 202.09 & 2 & 18 & 193 \\
Nr. exporters (HS10$\times$importer) & 1.98 & 1.63 & 1 & 1.59 & 3.35 \\
$HHI^{\text{suppliers,std}}$ & 0.35 & 0.31 & 0.06 & 0.24 & 0.98 \\
$HHI^{\text{suppliers}}$ & 0.83 & 0.17 & 0.59 & 0.86 & 1 \\
\bottomrule
\end{tabular}

\vspace{0.75cm}

\begin{tabular}{lccccc}
\multicolumn{6}{c}{(b) Importer Concentration} \\
\toprule
 & Mean & St. Dev. & p10 & p50 & p90 \\
\midrule
Nr. importers (HS10) & 59.11 & 139.79 & 2 & 16 & 152 \\
Nr. importers (HS10$\times$exporter) & 1.43 & 2.95 & 1 & 1.14 & 2.03 \\
$HHI^{\text{buyers,std}}$ & 0.39 & 0.31 & 0.07 & 0.28 & 1 \\
$HHI^{\text{buyers}}$ & 0.93 & 0.11 & 0.77 & 0.98 & 1 \\
\bottomrule
\end{tabular}
\par\end{centering}

\medskip
\centering
\parbox[t]{0.95\textwidth}{\scriptsize{}%
\textit{Note:} Summary statistics on exporter and importer concentration, calculated at the HS10 product-year level from 2011 to 2020. Panel (a) reports exporter concentration, while Panel (b) focuses on importer concentration. Exporters include firms from all countries, while importers are limited to Colombian firms. Network-based concentration measures ($HHI^{\text{suppliers}}$ and $HHI^{\text{buyers}}$) follow equations~\eqref{eq:HHI_suppliers} and~\eqref{eq:HHI_buyers}. Standard HHIs are computed using equations~\eqref{eq:HHI_suppliers_std} and~\eqref{eq:HHI_buyers_std}.}
\end{table}

First, although input markets involve many participants on both sides, firm-level sourcing and selling relationships are far more concentrated. The average HS10 product is imported from nearly 78 foreign exporters and purchased by around 59 Colombian firms. Yet the average Colombian importer relies on fewer than two foreign suppliers per product, and the typical exporter sells to fewer than 1.5 buyers. These sparse bilateral links help explain the high levels of concentration captured by network-based HHIs.

Second, importer concentration is more pronounced than exporter concentration. The average network-based HHI for importers is 0.93, compared to 0.83 for exporters, and importer HHIs are more skewed toward one. This partly reflects asymmetric coverage, as we observe all foreign exporters but only Colombian importers. However, the pattern persists even when we restrict to U.S.–Colombia trade and construct supplier and buyer shares using comparable single-country data for both sides. As shown in Appendix \ref{sec:USCOL}, importer concentration remains higher in this restricted case, though the gap is narrower.

Table~\ref{stats} also allows a comparison between network-based and standard industry-level concentration measures. On average, standard HHIs are substantially lower: 0.35 for exporters and 0.39 for importers, which is less than half the levels observed in the network-based measures. This discrepancy is partly mechanical, as standard HHIs define markets more broadly. Still, the magnitude of the gap suggests that conventional measures may significantly understate concentration in firm-to-firm trade.

\subsection{Time Trends in Aggregate Concentration}

\begin{figure}[t]
\begin{centering}
    \caption{Aggregate Concentration Trends, 2011--2020}
    \label{fig:concentrations}

    \begin{subfigure}{0.49\textwidth}
        \centering
        \includegraphics[width=1.1\textwidth]{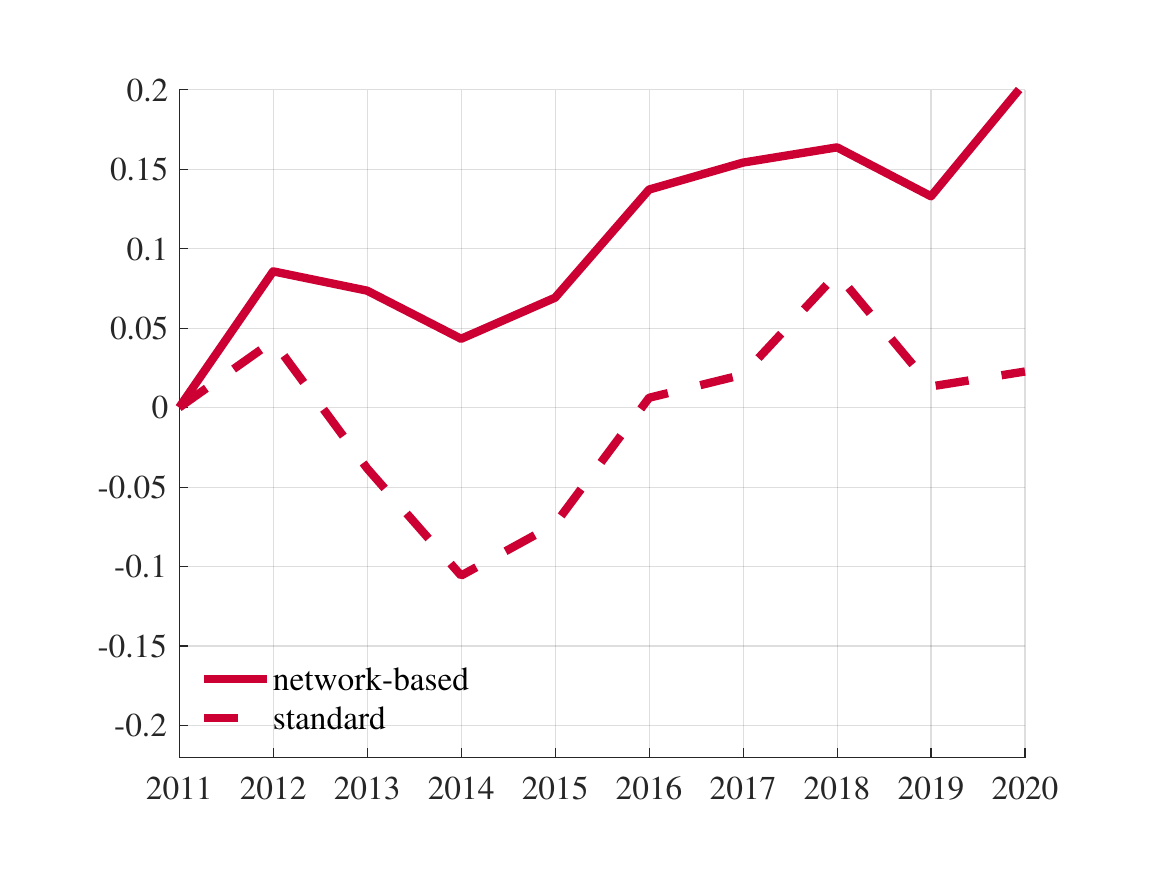}
        \caption{Exporter Concentration}
        \label{fig:supplier_concentration}
    \end{subfigure}%
    \hfill
    \begin{subfigure}{0.49\textwidth}
        \centering
        \includegraphics[width=1.1\textwidth]{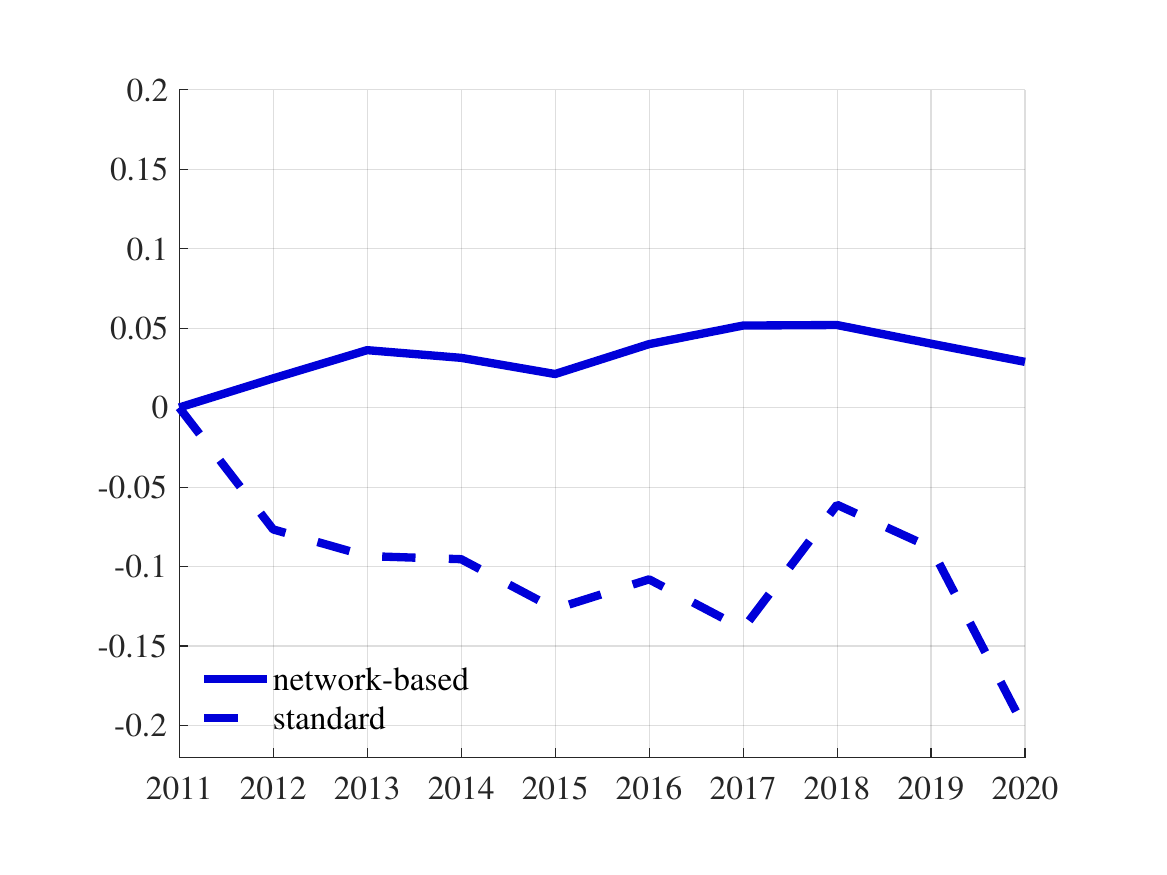}
        \caption{Importer Concentration}
        \label{fig:buyer_concentration}
    \end{subfigure}
    \par\end{centering}

    \textit{\scriptsize{}Notes: The solid line represents network-based concentration measures implied by our model with sticky firm-to-firm links, while the dashed line shows standard industry-wide HHIs. All series are normalized to their 2011 values.}{\scriptsize\par}
\end{figure}

These differences in levels naturally lead to the question of how the two measures compare in capturing changes in concentration over time. Figure~\ref{fig:concentrations} shows the evolution of aggregate concentration from 2011 to 2020, constructed using import-weighted averages across products, as defined in equation~\eqref{eq:agg_conc}. The solid lines show network-based concentration measures, while the dashed lines show the corresponding standard (industry-wide) HHIs.

Network-based measures indicate that exporter concentration increased by nearly 20 percent from 2011 to 2020, while importer concentration remained high, exhibiting a modest upward trend. In contrast, standard HHIs suggest flat or even declining concentration on both sides. These differences suggest that accounting for persistent firm-to-firm relationships can matter substantially, as the evolution of concentration, not just its level, may look very different depending on how markets are defined.

\section{Concentration and Aggregate Markups \label{sec:main_analysis}}

The influence of exporter and importer concentration on markups stems from the interplay of oligopoly and oligopsony forces, which are shaped by demand and supply elasticities, as well as bargaining power.  This section quantifies how concentration on both sides of the market interacts with these structural primitives to determine aggregate markups.

\subsection{Calibration and Estimation of Key Parameters\label{sec:Identification-and-estimation}}

We begin by outlining how the key structural parameters are calibrated or estimated.

\paragraph{Returns to Scale and Supply Elasticities}
Oligopsony power is governed by the parameter $\theta$, which captures the short-run returns to scale in exporters’ production. Specifically, $\theta$ determines the residual supply elasticity faced by each importer:
\begin{equation}
c^h_{q,ij} = \frac{d \ln c^h_{i}}{d \ln q^h_{ij}} = \frac{1 - \theta}{\theta} x^h_{ij}, \label{supply_elas}
\end{equation}
where $x^h_{ij}$ is the importer’s buyer share, as defined in equation~\eqref{x}.

We calibrate \(\theta\) using existing estimates of supply elasticities at the industry level. In U.S.\ manufacturing, \citet{boehm2022convex} report an inverse supply elasticity of 0.3 at median capacity utilization, and \citet{broda2008optimal} find similar values for the median HS4 product in the bottom tercile of goods, based on a sample of 15 countries between 1994 and 2003. Given that the median buyer share in our sample is close to one, equation~\eqref{supply_elas} implies a value of \(\theta = 0.75\) to match these estimates. We adopt this as our baseline calibration, viewing this choice as conservative.\footnote{\citet{broda2008optimal} report a wide range of inverse supply elasticities across four-digit HS goods, with values ranging from 0.3 to well above 100. This corresponds to values of \(\theta\) between 0 and 0.8. Our choice of \(\theta = 0.75\) lies at the upper end of this range and is therefore conservative. For comparison, \citet{alviarez2024twosided} estimate \(\theta = 0.5\) using U.S.\ import data in a closely related setting.}

\paragraph{Demand Elasticities}
The residual demand elasticity faced by exporter $i$ when selling to importer $j$, as defined in equation~\eqref{eq:mu_oligopoly}, depends on two key parameters: the within-industry elasticity of substitution, $\rho$, and the composite parameter $\eta$, which captures how sensitive importer $j$’s marginal cost is to changes in foreign input prices.

This elasticity varies with the supplier share $s_{ij}^h$: as $s_{ij}^h \to 0$, it converges to $\rho$; as $s_{ij}^h \to 1$, it approaches $\eta$. In Belgian data, \citet{kikkawa2022imperfect} estimate elasticities ranging from 2.5 to 7 in a framework similar to ours, with differentiated varieties and firm-to-firm trade. \citet{alviarez2024twosided} adopt baseline values of $\eta = 2.5$ and $\rho = 10$. In line with these studies, we set $\eta = 2.5$ and $\rho = 7$ in our baseline calibration.\footnote{These values are consistent with prior estimates of substitution elasticities between domestic and foreign intermediate inputs, which can be interpreted as measures of $\eta$ and provide a lower bound for $\rho$. For instance, \citet{Blaum2018gains} estimate an elasticity of 2.6, \citet{Halpern2015} report 4, and \citet{Costinot2014} use a value of 5. Our choices are therefore conservative.}

\paragraph{Bargaining Power}
Given parameters $\rho,\ \eta$, and $\theta$, we estimate the degree of importers’ bargaining power, $\phi$. Our data provides information on price variation across importers sourcing the same foreign supplier, product, and year, which allows us to identify $\phi$ directly following the strategy developed by \citet{alviarez2024twosided}. This is especially useful in our context, where there is limited prior evidence on the bargaining power of Colombian importers. Full details of the estimation procedure and results are provided in the companion paper and in Appendix~\ref{apndx:estimation}. To explore heterogeneity across input categories, we estimate $\phi$ separately across HS4 groupings.

\begin{figure}[t]
\caption{Importers' Bargaining Power Across Industries}
\label{fig:phi_estimates}
\begin{centering}
\includegraphics[width=0.6\paperwidth]{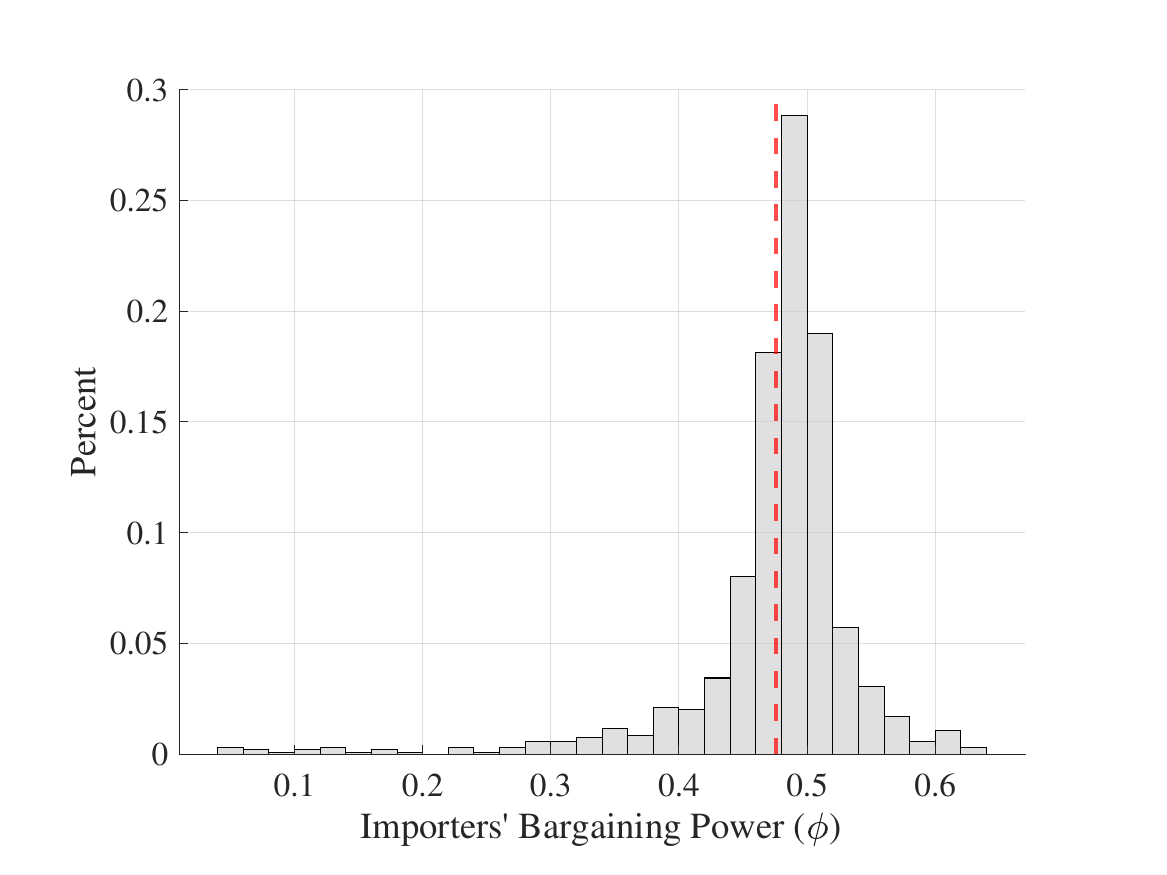}
\par\end{centering}
\textit{\scriptsize{}Notes: The figure shows estimates of importers’ bargaining power ($\phi$) across HS4 input categories. The vertical red dashed line indicates the mean $\phi$ across all HS4 inputs.}
\end{figure}

Figure~\ref{fig:phi_estimates} shows a histogram of the $\hat{\phi}$ estimates across HS4 input categories. Most estimates are highly significant at conventional levels. Importers’ bargaining power ranges from 0.05 to 0.64, with a mean and median around 0.5 and a standard deviation of 0.07, suggesting limited variation across input categories. On average, Colombian importers appear to have bargaining power comparable to that of their foreign suppliers.\footnote{These findings are consistent with related evidence from other settings. In U.S.\ data using the same estimation strategy, \citet{alviarez2024twosided} find that U.S.\ importers have four times the bargaining power of foreign exporters. Using a different methodology, \citet{atkin2024trade} show that Argentinian importers often hold substantial market power, with considerable heterogeneity across firms.}

We find that $\hat{\phi}$ is largely uncorrelated with concentration on either side of the market, whether measured using network-based or standard indices. The correlation between $\hat{\phi}$ and the import-weighted average importer concentration within an HS4 input group is $-0.05$ with network-based measures and $-0.10$ with standard ones. For exporter concentration, the corresponding correlations are $-0.10$ and $-0.11$. This lack of meaningful correlation suggests that bargaining power and concentration capture distinct dimensions of market power in international trade.

The parameter values also satisfy the conditions required for Assumptions \ref{A1} and \ref{A2} in the theoretical framework. Specifically, they imply a median value of $\lambda_{ij}$ across all pairs of 1, with a mean of 1.1, which in turn implies (via equation~\eqref{weight}) that $\hat{\omega}_{ij} \approx \hat{\phi}$, consistent with Assumption \ref{A1}.\footnote{Using the estimated parameter values and observed bilateral market shares $(s_{ij}, x_{ij})$, we compute $\lambda_{ij}$ and the corresponding markup $\mu_{ij}$ for each match according to equations~\eqref{weight} and~\eqref{eq:markup}, respectively.} The distribution of implied markups across matches is also tightly clustered around the mean, with an average of 1.2 and a standard deviation of 0.19. This pattern supports the use of the arithmetic mean as an approximation for the harmonic mean, as required by Assumption \ref{A2}.

\subsection{Aggregate Markup Trends} 
We now use observed changes in concentration indices, together with estimated parameters, to compute the implied evolution of aggregate markups on imported inputs, based on equation~\eqref{eq:agg_mkup_conc}. While the equation is expressed in levels, first-differencing yields changes in aggregate markups as a function of changes in buyer and supplier concentration, holding structural parameters fixed. It is important to note that this approach does not require the trade network to remain fixed over time. The theoretical framework accommodates an evolving network, as it only requires that firms treat the network as fixed within each individual (static) bargaining game.

\begin{figure}[t]
\caption{Evolution of Aggregate Markups in Input Trade, 2011--2020}
\label{fig:agg_mkp}
\begin{centering}
\includegraphics[width=0.6\paperwidth]{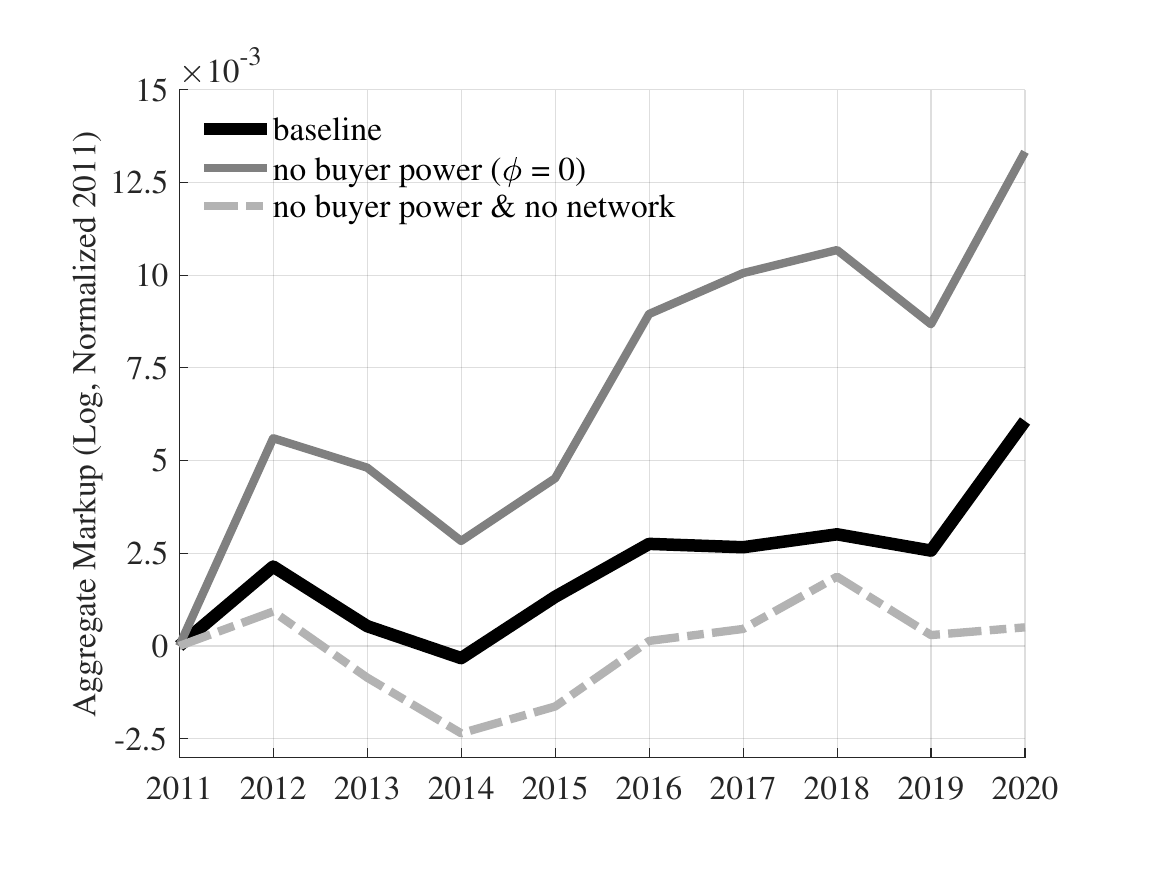}
\par\end{centering}
\textit{\scriptsize{}Notes: The solid black line shows the markup evolution implied by our baseline model, which accounts for bilateral market power. The solid red line eliminates importer bargaining power ($\phi = 0$), while the dashed red line further uses industry-wide exporter HHIs.}
\end{figure}

Figure~\ref{fig:agg_mkp} shows the results.  The baseline model (black line), which incorporates both network-based concentration and bilateral bargaining power, displays a moderate increase in aggregate markups of approximately 0.5\% over the decade. This rise is primarily driven by the sharp increase in network-based average exporter concentration, partially offset by a modest increase in countervailing oligopsony power, as reflected in the upward trend in the corresponding importer concentration shown in Figure~\ref{fig:concentrations}.

To quantify each mechanism, we consider two counterfactuals. The first (solid grey line) eliminates importer bargaining power by setting $\phi = 0$, holding market definitions fixed. In this case, markups rise by more than twice as much as in the baseline, highlighting the quantitative importance of buyer power in offsetting supplier concentration. 

The second counterfactual (dashed grey line) removes both importer bargaining power and network frictions, replacing the network-based exporter HHI with its standard sector-level counterpart. This case mirrors most empirical approaches that proxy market power using concentration. These approaches typically rely on standard oligopoly theory, where suppliers (exporters) set prices unilaterally and buyers (importers) act as passive price-takers. Markets are usually defined at the sectoral level, often using HS codes in trade data, with all firms within a sector treated as competitors.\footnote{For an application to international trade data, see, for example, \citet{Bonfiglioli2021}.} In this setting, markups closely track standard exporter concentration from Panel~(a) of Figure~\ref{fig:concentrations}, which shows little trend over time. As a result, the implied markup series remains nearly flat.

Taken together, these counterfactuals demonstrate that bilateral bargaining power and the definition of market structure through the trade network each play a critical (and conceptually distinct) role in determining the evolution of aggregate markups over time.

\section{Concluding Remarks}

This paper derives an analytical expression linking aggregate input markups to exporter and importer concentration in a model of firm-to-firm trade with two-sided market power and network rigidities. In this setting, two-sided concentration emerges as a sufficient statistic for aggregate markups, providing a theoretical foundation for using concentration measures as proxies for market power in global production networks.

A key insight of the paper is that, when trading relationships are persistent and competition is confined to existing links, standard industry-level market definitions can significantly understate true concentration and misrepresent its dynamics. We propose \emph{network-based} concentration indices that better capture competition in firm-to-firm trade. Although more data-intensive, these measures remain highly tractable and provide a transparent, model-consistent alternative to estimating markups.

Our theory implies that markets should be defined at the level of bilateral matches, consistent with the notion that firms compete only within their observed trading network. Whether firms outside this network exert latent competitive pressure remains an open empirical question, highlighting the need for further investigation in this area. 

Finally, although our focus is on input markets, the implications of input markups extend beyond the firm margin. A natural next step is to examine how supply chain concentration affects the pass-through of cost shocks, inflation dynamics, and welfare, with implications for the design of trade and competition policies.

\clearpage
\bibliographystyle{ecta}
\bibliography{library}
\clearpage

\appendix
\renewcommand{\thetable}{A.\arabic{table}} 
\renewcommand{\thefigure}{A.\arabic{figure}} 
\renewcommand{\theequation}{A.\arabic{equation}} 
\setcounter{table}{0}
\setcounter{figure}{0}
\setcounter{equation}{0}

\section*{Appendix}
\addcontentsline{toc}{section}{In-Paper Appendix}
\begingroup
\fontsize{10}{12}\selectfont
\setlength{\parindent}{0pt}
\setlength{\parskip}{1ex}

\section{Proof of Proposition \ref{prop:1} \label{subsec:Derivation-of-approximation}}

Leveraging Assumptions \ref{A1} and \ref{A2} in the main text gives the following approximation for aggregate industry markups:
\[
\mu \approx (1 - \phi) \sum_{i}\sum_{j} \iota_{ij} \mu_{ij}^{\text{oligopoly}} 
+ \phi \sum_{i}\sum_{j} \iota_{ij} \mu_{ij}^{\text{oligopsony}}.
\]
where $\iota_{ij} \equiv \frac{p_{ij} q_{ij}}{\sum_{i} \sum_{j} p_{ij} q_{ij}}$.

For the oligopoly term, we follow \cite{Grassi2018} and expand $\mu_{ij}^{\text{oligopoly}}$ in powers of $s_{ij}$:
\begin{align*}
\mu_{ij}^{\text{oligopoly}} &= \frac{1}{1 - \varepsilon_{ij}^{-1}} 
= \frac{1}{1 - \frac{1}{\rho - (\rho - \eta)s_{ij}}}
= \frac{1}{1 - \frac{1}{\rho} \sum_{m=0}^{\infty} \left( \frac{\rho - \eta}{\rho} \right)^m s_{ij}^m}.
\end{align*}
Retaining terms up to second order and neglecting higher-order terms, we obtain:
\[
\sum_{ij} \iota_{ij} \mu_{ij}^{\text{oligopoly}} 
\approx \frac{\rho}{\rho - 1} + \frac{\rho - \eta}{(\rho - 1)^2} HHI^{\text{suppliers}},
\]
where we defined the following weighted supplier concentration index: \(HHI^{\text{suppliers}} \equiv \sum_{j} \varphi_j \sum_{i} s_{ij}^2\), with \(\varphi_j = \frac{\sum_i p_{ij} q_{ij}}{\sum_{ij} p_{ij} q_{ij}}\).

For the oligopsony term, we approximate:
\[
\mu_{ij}^{\text{oligopsony}} \approx 1 + \frac{\theta - 1}{2\theta} x_{ij}.
\]
Although this is a local expansion around \( x_{ij} = 0 \), it provides a good global approximation for  values of \( \theta \in (0.4, 0.75) \). It outperforms the alternative expansion around \( x_{ij} = 1 \) in terms of fit.

Applying this approximation and ignoring higher-order terms yields:
\[
\sum_{ij} \iota_{ij} \mu_{ij}^{\text{oligopsony}} 
\approx 1 + \frac{\theta - 1}{2\theta} HHI^{\text{buyers}},
\]
where we defined the following buyer-side concentration index: \(HHI^{\text{buyers}} \equiv \sum_{i} \varphi_i \sum_{j} x_{ij}^r x_{ij},\) where \(x_{ij}^r = \frac{p_{ij} q_{ij}}{\sum_{j} p_{ij} q_{ij}}, \quad
\varphi_i = \frac{\sum_{j} p_{ij} q_{ij}}{\sum_{ij} p_{ij} q_{ij}}. \)

Combining both terms, we obtain the final approximation for the aggregate markup:
\begin{align*}
\mu 
&\approx (1 - \phi) \frac{\rho}{\rho - 1} + \phi 
+ (1 - \phi) \frac{\rho - \eta}{(\rho - 1)^2} HHI^{\text{suppliers}} 
- \phi \frac{1 - \theta}{2\theta} HHI^{\text{buyers}}.
\end{align*}

This expression decomposes the aggregate markup into a baseline level (determined by structural elasticities), an upward correction reflecting supplier concentration (oligopoly), and a downward correction reflecting buyer concentration (oligopsony). This completes the proof of Proposition~\ref{prop:1}.

\clearpage

\renewcommand{\thetable}{C.\arabic{table}} 
\renewcommand{\thefigure}{C.\arabic{figure}} 
\renewcommand{\theequation}{C.\arabic{equation}} 
\setcounter{table}{0}
\setcounter{figure}{0}
\setcounter{equation}{0}

\section*{Online Appendix}
\addcontentsline{toc}{section}{Online Appendix}

\section{Input Trade, Concentration, and Consumer Prices}
\label{app:consumer-prices}

We show an example of how the theoretical framework can be extended to link markups in input trade to consumer prices through their impact on firms’ cost structures.

Given the production function for importing firm~$j$ in equation~\eqref{eq:q_j}, the unit cost incurred by firm~$j$ to produce its final good is:
\begin{equation}
    c_j = \frac{1}{\varphi_j} \left( \frac{p^d}{1 - \gamma} \right)^{1 - \gamma} \left( \frac{p^f_j}{\gamma} \right)^\gamma,
\end{equation}
where $\varphi_j$ is firm $j$’s productivity, $p^d$ is the (common) price of the domestic input (e.g., labor), and $p^f_j$ is the price of the imported input, given by
\begin{equation}
    p^f_j = \left( \sum_i p_{ij}^{1 - \rho} \right)^{\frac{1}{1 - \rho}}.
\end{equation}
This formulation links firms’ unit costs directly to import prices and, by extension, to the bilateral markups embedded in those prices.

To examine the downstream implications, we assume that the outputs of importing firms are aggregated into a final consumption bundle using CES preferences with elasticity of substitution $\nu$.\footnote{For simplicity, we assume that all downstream firms are importers. Alternatively, one could assume a Cobb--Douglas aggregator over the output of importing and non-importing firms. In that case, the expressions below would be scaled by the share of importers in total consumption, but the qualitative results would remain unchanged.} Firms compete in monopolistic competition and set prices as a constant markup over marginal cost, charging $p_j = \frac{\nu}{\nu - 1} c_j$.

The resulting price index for the final output bundle is:
\begin{equation}
    P = \left( \sum_j \zeta_j^{\nu} \left( \frac{\nu}{\nu - 1} c_j \right)^{1 - \nu} \right)^{\frac{1}{1 - \nu}},
    \label{eq:PF}
\end{equation}
where $\zeta_j$ is a firm-specific demand shifter and $c_j$ is defined above.

To assess how input markups may affect consumer prices, we define a distortion index on final goods as the ratio of the actual price index to a counterfactual one $\tilde{P}$ in which all import markups are absent (i.e., $\mu_{ij} = 1$ for all $i,j$). This approach resembles classic ``conduct" methods used to quantify market power distortions, such as in \citet{nocke2025concentration}.\footnote{See also the discussion in \citet{bresnahan1989empirical}.} We derive:
\begin{equation}
    \hat{P} \equiv  \left( \frac{\tilde{P}}{P} \right)^{-1} = \left( \sum_j \omega_j \left( \sum_i s_{ij} \left( \frac{1}{\mu_{ij}} \right)^{1 - \rho} \right)^{\frac{1 - \nu}{1 - \rho} \gamma} \right)^{\frac{-1}{1 - \nu}},
    \label{eq:mu_F_exact}
\end{equation}
where $\omega_j \equiv \frac{\zeta_j^{\nu} p_j^{1 - \nu}}{\sum_k \zeta_k^{\nu} p_k^{1 - \nu}}$ is importer $j$’s share in final output, and $s_{ij}$ is exporter $i$’s supplier share. 

Following derivations analogous to those in Proposition~\ref{prop:1}, we obtain the following approximation:
\begin{equation}
    \hat{P} \approx (1 - \gamma) \left( \frac{\rho}{\rho - 1} \right)^{\gamma} 
    + \left( \frac{\rho}{\rho - 1} \right)^{-(1 - \gamma)} \gamma \underbrace{\sum_i \sum_j \iota_{ij} \mu_{ij}}_{\mu},
    \label{eq:mu_F_approx}
\end{equation}
where $\iota_{ij}$ denotes the share of importer $j$’s total spending allocated to supplier $i$. The final term corresponds to the aggregate imported-input markup, denoted $\mu$, which is analyzed in the main text and shown in Proposition~\ref{prop:1} to be governed by two-sided concentration in input markets.

This expression shows that markups in input trade affect the level of final consumer prices via their effect on importers’ unit costs. When $\gamma = 1$, that is, when firms rely entirely on imported inputs, equation~\eqref{eq:mu_F_approx} collapses to the aggregate markup expression in Section~\ref{subsec:Aggregate-Markups}. More generally, for any $\gamma > 0$, the bilateral markups $\mu_{ij}$ affect the final price index, with the effect being proportional to $\gamma$.

While this analysis rests on strong assumptions, including full pass-through, Cobb-Douglas technology, and common sourcing strategies, it shows that concentration in input trade may serve as a sufficient statistic for the distortionary effects of input markups on consumer welfare.

\newpage 

\section{Additional Empirical Results\label{sec:Additional-Tables-and}}

\subsection{Colombian Imports}

Table \ref{tab:Colombian-Imports:-Statistics} validates the Colombian customs records used in our analysis against official statistics from UN Comtrade. For each year from 2011 to 2020, we report total imports in million USD from both sources, along with the number of transactions, firm-to-firm trade pairs, unique importers, and unique foreign suppliers in the customs data. Our dataset closely tracks aggregate import values reported by Comtrade. 

\begin{table}[h]
\centering
\caption{Colombian Imports: Validation and Coverage \label{tab:Colombian-Imports:-Statistics}}

\begin{tabular}{ccccccc}
\toprule
Year & \multicolumn{2}{c}{Imports (Million USD)} & Transactions & Pairs & Importers & Suppliers \\
 & Customs & Comtrade & & & (Buyers) & (Suppliers) \\
\midrule
2011 & 60{,}467 & ---     & 2{,}749{,}052 & 164{,}324 & 32{,}541 & 81{,}402 \\
2012 & 58{,}767 & 56{,}833 & 2{,}951{,}104 & 172{,}178 & 35{,}240 & 85{,}616 \\
2013 & 53{,}238 & 58{,}443 & 2{,}857{,}520 & 162{,}931 & 33{,}005 & 83{,}798 \\
2014 & 46{,}804 & 62{,}939 & 2{,}434{,}251 & 157{,}281 & 32{,}187 & 81{,}942 \\
2015 & 52{,}064 & 54{,}035 & 3{,}036{,}051 & 164{,}473 & 32{,}399 & 85{,}807 \\
2016 & 28{,}816 & 44{,}831 & 1{,}915{,}076 & 134{,}450 & 28{,}922 & 73{,}621 \\
2017 & 26{,}892 & 46{,}050 & 2{,}015{,}542 & 126{,}379 & 27{,}500 & 70{,}392 \\
2018 & 29{,}041 & 51{,}230 & 1{,}936{,}597 & 123{,}599 & 26{,}973 & 68{,}630 \\
2019 & 53{,}793 & 52{,}263 & 3{,}665{,}783 & 164{,}863 & 32{,}429 & 85{,}173 \\
2020 & 27{,}294 & 43{,}487 & 1{,}835{,}985 & 115{,}656 & 25{,}867 & 64{,}560 \\
\midrule
Mean & 43{,}578 & --- & 2{,}492{,}217 & 147{,}599 & 30{,}416 & 77{,}466 \\
\bottomrule
\end{tabular}

\vspace{0.25cm}
\parbox[t]{0.97\textwidth}{\scriptsize
\textit{Note:} The table compares transaction-level Colombian customs data to official import statistics from UN Comtrade (column 3). Columns 4--7 summarize the scope and granularity of the customs records, including the number of annual transactions, firm-to-firm trade pairs, and unique importing and exporting firms. Values are aggregated annually over the period 2011–2020. Discrepancies between customs and Comtrade values reflect differences in reporting scope and methodology.}
\end{table}

\pagebreak
\subsection{U.S.-Colombia Trade \label{sec:USCOL}}

Table~\ref{stats_uscol} replicates Table~\ref{stats}, restricting the sample to trade between Colombia and the United States. Relative to the full sample, the Colombia–US trade network is notably more concentrated, as expected. The average number of U.S. exporters per HS10 product drops from 78 in the full sample to just 29, while the average number of Colombian importers per product falls from 59 to 24. This sparser network translates into higher concentration indices: the average network-based exporter concentration rises from 0.83 to 0.90, and the buyer-side HHI from 0.93 to 0.94. 

\begin{table}[h]
\caption{Exporter and Importer Concentration (Colombia--US Trade): Summary Statistics\label{stats_uscol}}

\begin{centering}
\small
\begin{tabular}{lccccc}
\multicolumn{6}{c}{(a) Exporter Concentration} \\
\toprule
 & Mean & St. Dev. & p10 & p50 & p90 \\
\midrule
Nr. U.S. exporters (HS10) & 28.88 & 76.94 & 1.00 & 7.00 & 68.00 \\
Nr. U.S. exporters (HS10$\times$importer){\color{white}mbian} & 1.51 & 1.16 & 1.00 & 1.24 & 2.27 \\
$HHI^{\text{suppliers,std}}$ & 0.51 & 0.32 & 0.12 & 0.45 & 1.00 \\
$HHI^{\text{suppliers}}$ & 0.90 & 0.14 & 0.69 & 0.98 & 1.00 \\
\bottomrule
\end{tabular}

\vspace{0.75cm}

\begin{tabular}{lccccc}
\multicolumn{6}{c}{(b) Importer Concentration} \\
\toprule
 & Mean & St. Dev. & p10 & p50 & p90 \\
\midrule
Nr. Colombian importers (HS10) & 24.34 & 59.06 & 1.00 & 6.00 & 59.00 \\
Nr. Colombian importers (HS10$\times$exporter) & 1.33 & 1.08 & 1.00 & 1.00 & 1.82 \\
$HHI^{\text{buyers,std}}$ & 0.53 & 0.33 & 0.12 & 0.48 & 1.00 \\
$HHI^{\text{buyers}}$ & 0.94 & 0.12 & 0.80 & 1.00 & 1.00 \\
\bottomrule
\end{tabular}
\par\end{centering}

\medskip
\centering
\parbox[t]{0.95\textwidth}{\scriptsize{}%
\textit{Note:} Summary statistics on exporter and importer concentration, calculated at the HS10 product-year level from 2011 to 2020, restricted to Colombia--US trade. Panel (a) reports exporter concentration, while Panel (b) focuses on importer concentration. Network-based concentration measures ($HHI^{\text{suppliers}}$ and $HHI^{\text{buyers}}$) follow equations~\eqref{eq:HHI_suppliers} and~\eqref{eq:HHI_buyers}. Standard HHIs are computed using equations~\eqref{eq:HHI_suppliers_std} and~\eqref{eq:HHI_buyers_std}.}
\end{table}

\pagebreak

\section{Estimation of the Bargaining Parameter {$\phi$}}\label{apndx:estimation}

We denote the log bilateral price of product $h$ traded between exporter $i$ and importer $j$ in year $t$ as:
\begin{equation}
\ln p_{ijt}^{h} = \ln \mu(\phi, \theta; \Omega_{ijt}) + \ln c_{it}^{h} + k_{ijt}^{h},
\end{equation}
where $\Omega_{ijt}$ is the information set available to the exporter-importer pair $(i,j)$ during their negotiation. This set includes the supplier and buyer shares, $s_{ijt}^{h}$ and $x_{ijt}^{h}$, as well as the calibrated parameters $(\nu, \gamma, \rho, \theta)$. From the functional form of the markup in equation~\eqref{eq:markup}, it follows that conditional on $\Omega_{ijt}$, the bilateral markup depends only on the model primitive $\phi$, i.e., $\mu_{ij} = \mu(\phi; \Omega_{ijt})$.

The log price can therefore be decomposed into the log markup and the log marginal cost (inclusive of duties). We express the marginal cost as the sum of an exporter-specific term, $\ln c_{it}^{h}$, which is common across all importers to exporter $i$, and an idiosyncratic term $k_{ijt}^{h}$. This latter term is assumed to be mean-zero, i.i.d. across buyers of the same supplier, and captures unobserved heterogeneity in marginal costs due to factors not explicitly modeled, such as quality differentiation or input customization.

Under this specification, the expected difference in exporter $i$'s marginal cost between two importers $j$ and $\ell$, conditional on the joint information set $\boldsymbol{\Omega}_{ij\ell t} \equiv (\Omega_{ijt}, \Omega_{i\ell t})$, is zero:
\begin{equation}
\mathbb{E}_u \left[ k_{ijt}^{h} - k_{i\ell t}^{h} \mid \boldsymbol{\Omega}_{ij\ell t} \right] = 0.
\end{equation}
Taking the difference in the expected log prices charged by exporter $i$ to importers $j$ and $\ell$, we obtain the following moment condition:
\begin{equation}
\label{eq:moment-cond}
g(\phi, \theta; \boldsymbol{\Omega}_{ij\ell t}) \equiv \mathbb{E}_u \left[ \ln p_{ijt}^{h} - \ln p_{i\ell t}^{h} - \left( \ln \mu(\phi; \Omega_{ijt}) - \ln \mu(\phi; \Omega_{i\ell t}) \right) \mid \boldsymbol{\Omega}_{ij\ell t} \right] = 0, \quad \forall\ i,j,\ell,t.
\end{equation}
The identification of $\phi$, as discussed in \citet{alviarez2024twosided}, is based on this condition.

We estimate equation~\eqref{eq:moment-cond} using the generalized method of moments (GMM):
\begin{equation}
\min_{\phi, \theta} \, \mathbf{g}(\phi)' \, \mathbf{Z}' \, \mathbf{W} \, \mathbf{Z} \, \mathbf{g} (\phi),
\end{equation}
where $\mathbf{g}(\phi)$ stacks all moment conditions from equation~\eqref{eq:moment-cond} across all $(i,j,\ell,t)$ quadruples, $\mathbf{Z}$ denotes the matrix of instruments (needed to account for endogeneity between $k$ and bilateral shares), and $\mathbf{W}$ is the optimal weighting matrix.
\endgroup

\end{document}